\documentclass[10pt,conference]{IEEEtran}
\usepackage{epsfig}
\usepackage{amsmath}
\usepackage{amsfonts}
\usepackage{amssymb}
\usepackage[all]{xy}
\usepackage{xcolor}
\usepackage{comment}
\usepackage{perso}
\usepackage{graphicx}
 \graphicspath{{eps/}} \DeclareGraphicsExtensions{.eps}

\newtheorem{theo}{Theorem}
\newtheorem{prop}[theo]{Proposition}

\newtheorem{coro}[theo]{Corollary}

\newcommand{\pth}{p_\txtscript{th}}
\newcommand{\dvmax}{d_v}
\newcommand{\dcmax}{d_c}
\newcommand{\Gr}{\mbox{\rm Gr}}
\newcommand{\muave}{\mu_m}

\title{Non binary LDPC codes over the binary erasure channel: density evolution analysis}
\author{Valentin~Savin, CEA-LETI, MINATEC, Grenoble, France, valentin.savin@cea.fr}
\date{}
\begin{document}
\maketitle

%%-------------------------------------------------------------------------------
\begin{abstract}
\renewcommand{\thefootnote}{} \footnote{This work has been supported by the French ANR grant
N° 2006 TCOM 019 (CAPRI-FEC project)}\setcounter{footnote}{0}
 In this paper we present a thorough analysis of non binary LDPC codes
over the binary erasure channel. First, the decoding of non binary
LDPC codes is investigated. The proposed algorithm performs ``on-the
fly'' decoding, {\em i.e.} it starts decoding as soon as the first
symbols are received, which generalizes the erasure decoding of
binary LDPC codes. Next, we evaluate the asymptotical performance of
ensembles of non binary LDPC codes, by using the density evolution
method. Density evolution equations are derived by taking into
consideration both the irregularity of the bipartite graph and {\em
the probability distribution of the graph edge labels}. Finally,
infinite-length performance of some ensembles of non binary LDPC
codes for different edge label distributions are shown.

\end{abstract}

% \begin{keywords}
% \end{keywords}

%%-------------------------------------------------------------------------------
%%-------------------------------------------------------------------------------
\section{Introduction}

Data loss recovery -- for instance, for content distribution
applications or for distributed storage systems -- is widely
addressed using FEC (Forward Error Correction) techniques based on
error correcting codes. These codes are dealing with erasure
channels, {\em i.e.} a channel that either transmits the data unit
correctly (without error) or erases it completely. In the case of
content distribution applications, the potential physical layer CRC,
or physical layer FEC codes, or transport level UDP checksums, may
lead a receiver to discard erroneous data units. For distributed
storage systems, data loss may be due to broken servers,
Denial-of-Service (DoS) attacks, etc.

The performance of error correcting codes over erasure channels can
be analyzed precisely, and a flurry of research papers have already
addressed this issue.
%For small to medium codeword length, {\em
%Maximum-Distance Separable} (MDS) codes achieve the channel
%capacity. However, for large block lengths, their decoding becomes
%untractable, and thus iteratively decoded graph-based codes
%constitute the main alternative.
Low-density parity-check (LDPC)
codes \cite{gall_phd}, \cite{gall-monograph} with iterative decoding
\cite{zyablov} proved to perform very close to the channel capacity
with reasonable complexity \cite{richardson_design_2001}
\cite{luby2001eec}. Moreover, ``rateless'' codes that are capable of
generating an infinite sequence of repair symbols were proposed in
\cite{luby2002lc} \cite{shokrollahi2006rc}. LDPC codes were
generalized by Tanner \cite{Tann}, by introducing the sparse graph
representation and replacing the Single Parity Check (SPC)
constraint nodes with error correcting block codes. Nowadays, these
codes are known as GLDPC codes and were recently investigated for
the BEC \cite{paolini:agl}, \cite{paolini:gsc},
 \cite{mila-foss-gldpc}. Over the past few years there has been an increased interest in
 non binary LDPC codes due to their enhanced correction capacity,
 but at this time only few works are dealing with the BEC \cite{rathi:det},\cite{rathi:cen}.
In this paper we give a thorough analysis of non binary LDPC codes
over the BEC. The paper is organized as follows: in Section
\ref{non_binary_ldpc} we review some background on the construction
of non binary LDPC codes. The decoding of non binary LDPC codes over
the BEC is addressed in Section \ref{decoding_bec}. In Section
\ref{density_evolution_bec} we derive the density evolution
equations taking into consideration both the irregularity of the
bipartite graph and the probability distribution of the graph edge
labels. Thresholds of some ensembles of non binary LDPC codes for
different edge label distributions are shown in Section
\ref{thresholds}.

\section{Non binary LDPC codes} \label{non_binary_ldpc}
We denote by $\gf_q$ the Galois field with $q$ elements. For
practical reasons, we will assume that $q$ is a power of $2$, even
if this condition is not always necessary. Thus, we set $q = 2^p$,
where $p$ is the vector space dimension of $\gf_q$ over $\gf_2$
(each time we refer to $\gf_q$ as a vector space, we consider its
$\gf_2$-vector space structure). We fix once for all an isomorphism
of $\gf_2$-vector spaces:

%We identify $\gf_{q} = \{0,1,\dots,q-1\}$ and call these elements  $\gf_q$-{\it symbols}, or simply {\em symbols}, in order to be
%distinguished from ordinary integers.
%\begin{equation}
%    \label{identify}
%  \begin{array}{ccc}
%  \gf_2^p & \rightarrow & \gf_q \\
%  (b_0,\dots,b_{p-1}) & \mapsto & \displaystyle\sum_{i=0}^{p-1}b_i2^i
%  \end{array}
%\end{equation}
%and we say that $(b_0,\dots,b_{p-1})$ are the constituent bits of the symbol $s$ if $s = \displaystyle\sum_{i=0}^{p-1}b_i2^i$.

\begin{equation}
    \label{identify}
    \gf_2^p \stackrel{\sim}{\rightarrow} \gf_q
\end{equation}
and we say that $(b_0,\dots,b_{p-1})\in\gf_2^p$ are the constituent
bits of the symbol $s\in\gf_q$, if they correspond to each other by
the above isomorphism.

Let $\mathbb{L}$ be a multiplicative group acting on the vector space $\gf_q$. For instance, we may have:
\begin{itemize}
  \item $\mathbb{L} = \gf_q^{*}$, acting on $\gf_q$ via the internal field multiplication;
  \item $\mathbb{L} = \mbf{M}_p^{*}(\gf_2)$, the multiplicative group of invertible $p\times p$ matrices, acting on $\gf_q$ via the
  isomorphism $\gf_2^p \stackrel{\sim}{\rightarrow} \gf_q$ from (\ref{identify}).
\end{itemize}
The action of $\mathbb{L}$ on $\gf_q$ will always be denoted multiplicatively, that is:
%\begin{equation}
%  \label{action}
%  \begin{array}{ccc}
%    \mathbb{L} \times \gf_q & \rightarrow & \gf_q \\
%    (h, s) &\mapsto & hs
%  \end{array}
%\end{equation}
\begin{equation}
  \label{action}
    \mathbb{L} \times \gf_q  \rightarrow  \gf_q:\ \  (h, s) \mapsto  hs
\end{equation}

For any matrix $H \in \mbf{M}_{M,N}(\mathbb{L})$ one can define a
 code:
\begin{eqnarray}
  \mcl{C} \hspace{-2mm}&=\hspace{-2mm}& \ker(H) \\
     &=\hspace{-2mm}& \{ (s_1,\dots,s_N)\mid \sum_{n=1}^N h_{m,n}s_n = 0,\ \forall m = 1,\dots,M
     \}\nonumber
\end{eqnarray}
If $\mathbb{L} = \gf_q^{*}$ acting on $\gf_q$ via the internal field
multiplication, then $\mcl{C}$ is a $\gf_q$-linear code, but this
does not happen for general $\mathbb{L}$.

The Tanner graph associated with the code $\mcl{C}$, denoted by
${\cal H}$, consists of $N$ {\it variable nodes} and $M$ {\it check
nodes} representing the $N$ columns and the $M$ lines of the matrix
$H$. A variable node and a check node are connected by an edge if
the corresponding element of matrix $H$ is not zero. Each edge of
the graph is labeled by the corresponding non zero element of $H$.
Thus, from now on, we refer to the elements of $\mbb{L}$ as {\em
labels}. We also denote ${\cal H}(n)$ the set of check nodes
connected to a given variable node $n \in \{ 1, 2, \dots, N \}$, and
by ${\cal H}(m)$ the set of variable nodes connected to a given
check node $m \in \{ 1, 2, \dots, M \}$.

%The following notation will be used throughout the paper with
%respect to the Tanner graph:
%\begin{list}{$\bullet$}{\itemindent 0mm \leftmargin\parindent}
%\begin{itemize}
%\item ${\cal H}$, the Tanner graph of the code.
%\item $n \in \{ 1, 2, \dots, N \}$ a {\it variable node} of ${\cal H}$.
%\item $m \in \{ 1, 2, \dots, M \}$ a {\it check node} of ${\cal H}$.
%\item $(m,n)$ denotes an {\em oriented edge} connecting the {\em outgoing} check node $m$ to the {\em ingoing} variable
%node $n$.
%\item $(n,m)$ denotes an {\em oriented edge} connecting the {\em outgoing} variable
%node $n$ to the {\em ingoing} check node $m$.
%\item ${\cal H}(n)$, set of check nodes connected to a given variable node $n$.
%\item ${\cal H}(m)$, set of variable nodes connected to a given check node $m$.
%\end{itemize}
%\end{list}

\subsection{The binary image of a non binary code}
Any sequence $(s_1,\dots,s_N)\in\gf_q^N$ may be mapped into a binary
sequence of length $Np$ via the isomorphism of (\ref{identify}). The
binary sequences associated with the codewords
$(s_1,\dots,s_N)\in\mcl{C}$ constitute a linear binary code
$\mcl{C}_\txtscript{bin} \subseteq \gf_2^{Np}$, which is called the
{\em binary image} of $\mcl{C}$. Moreover, the action (\ref{action})
of the multiplicative label group $\mbb{L}$ on $\gf_q$  induces a
group morphism from $\mbb{L}$ into the group of vector space
endomorphisms $\mcl{L}_{\gf_2}(\gf_q,\gf_q)$, and identifying
$\gf_q$ and $\gf_2^p$ via (\ref{identify}), we get a morphism:
\begin{equation}
  \mbb{L} \rightarrow \mcl{L}_{\gf_2}(\gf_q,\gf_q) \stackrel{\sim}{\rightarrow} \mcl{L}_{\gf_2}(\gf_2^p,\gf_2^p) = \mbf{M}_{p}(\gf_2)
\end{equation}
Replacing each coefficient of the matrix
$H\in\mbf{M}_{M,N}(\mbb{L})$ with its image under the above
morphism, we obtain a binary matrix
$H_\txtscript{bin}\in\mbf{M}_{Mp,Np}(\gf_2)$, which is simply the
parity check matrix of the binary code $\mcl{C}_\txtscript{bin}$.
While the encoding may be performed using either the non binary code
or its binary image, the iterative decoding of a non binary code on
its binary image generally yields very poor performance.

\section{Decoding non binary LDPC codes}\label{decoding_bec}

For general channels, several decoding algorithms for non binary
LDPC codes were proposed in the literature \cite{wiberg},
\cite{declercq_extended_2005}, \cite{savin_min_max}. Because of the
BEC specificity, these algorithms are all equivalent over the BEC,
and they can be described in a slightly different manner, as
presented below.

\subsection{Decoding over the BEC}
In this section we assume that a non binary LDPC code is used over
BEC($\epsilon$) -- the binary erasure channel with erasure
probability $\epsilon$. Thus, the length $N$ sequence of encoded
$\gf_{q}$-symbols is mapped into the corresponding binary sequence
of length $Np$, which is transmitted over the BEC, each bit from the
binary sequence being erased with probability $\epsilon$. We say
that a $\gf_{q}$-symbol is :
\begin{itemize}
\item {\em received}, if all of its constituent bits are received;
\item {\em erased}, if all of its constituent bits are erased by the channel;
\item {\em partially erased}, if some of its constituent bits are erased by the channel and some others are received.
\end{itemize}
At the receiver part, the received bits are used to reconstruct the
corresponding $\gf_{q}$-symbols. The reconstruction may be complete,
partial, or lacking, according to whenever the corresponding symbol
is received, partially erased, or erased.

Let $n$ be a variable node of the Tanner graph and $s\in\gf_q$. We
say that the symbol $s$ is {\em eligible} at the variable node $n$,
if the probability of the $n^\txtscript{th}$ transmitted symbol
being $s$ is non zero. Tacking into consideration the channel
output, the {\em a priori set of eligible symbols}, denoted by
$\msr{E}_n$, consists of the symbols that fit with the received
constituent bits (if any) of the $n^\txtscript{th}$ transmitted
symbol. Thus :
\begin{itemize}
\item $\msr{E}_n = \gf_{q}$, if the symbol is erased,
\item $\msr{E}_n \varsubsetneq \gf_{q}$, if the symbol is partially erased,
\item $\card(\msr{E}_n) = 1$, if the symbol is received.
\end{itemize}

 These sets constitute the {\em a priori information} of the decoder. They are iteratively updated by exchanging
extrinsic messages between variable and check nodes in the graph. Each message is a subset of $\gf_{q}$, representing a set of eligible
symbols. Precisely, the message sent by a graph node on an outgoing edge is a set of eligible symbols, which is computed according to
messages received by the same node on the incoming edges. We use the following notation:
\begin{itemize}
\item $\msr{A}_{m,n}$ the set of eligible symbols sent by the variable node $n$ to the check node $m$;
\item $\msr{B}_{m,n}$ the set of eligible symbols sent by the check node $m$ to the variable node $n$.
\end{itemize}
Finally, if $\msr{S}, \msr{S}_1, \msr{S}_2\subseteq\gf_q$ and $h\in\mbb{L}$ we define:
 $$\begin{array}{rcl}
   h\msr{S} & = & \{ hs \mid s\in\msr{S} \} \\
   \msr{S}_1 + \msr{S}_2 & = & \{ s_1 + s_2 \mid s_1\in\msr{S}_1, s_2\in\msr{S}_2 \}
 \end{array}$$

The iterative decoder for the BEC can be expressed as follows:

\medskip\noindent\underline{\bf Initialization step}

\begin{itemize}
\item variable-to-check messages initialization

$\msr{A}_{m,n} = \msr{E}_n$
\end{itemize}

\medskip\noindent\underline{\bf Iteration step}

\begin{itemize}
\item check-to-variable messages

$\msr{B}_{m,n} = \displaystyle\sum_{n'\in\mcl{H}(m)\setminus\{n\}} h_{m,n'} \msr{A}_{m,n'}$

\bigskip
\item variable-to-check messages

$\msr{A}_{m,n} = \displaystyle\msr{E}_n \cap\left(\bigcap_{m'\in\mcl{H}(n)\setminus\{m\}} h_{m',n}^{-1} \msr{B}_{m',n}\right)$

\bigskip
\item a posteriori sets of eligible symbols

$\overline{\msr{E}}_{n} = \displaystyle\msr{E}_n \cap\left(\bigcap_{m\in\mcl{H}(n)} h_{m,n}^{-1} \msr{B}_{m,n}\right)$
\end{itemize}
The decoder stops when all the a posteriori sets of eligible symbols
$\overline{\msr{E}}_{n}$ are of cardinality $1$, or when a maximum
number of iterations is reached. It is important to note that any
set of eligible symbols ($\msr{E}_n, \msr{A}_{m,n}, \msr{B}_{m,n}$,
or $\overline{\msr{E}}_{n}$) is a $\gf_{2}$-affine sub-space of
$\gf_{q}$; in particular, its cardinal is a power of $2$.

\subsection{Minimum-delay decoding} \label{minimum_delay}
In this section we propose a {\em minimum-delay} decoding algorithm
over the BEC, in the sense that the decoding starts since the
reception of the first bits, which is suited for Upper-Layer Forward
Error Correction (UL-FEC).
%For binary LDPC codes, a minimum-delay
%decoding, called {\em erasure decoding}, was proposed in
%\cite{luby2001eec}, \cite{luby1997plr}. For binary codes, the
%minimum-delay decoding removes bit nodes and incident edges from the
%graph as follows: each time a new bit is received from the channel,
%its value is added to the neighbor check nodes, then the
%corresponding bit node and incident edges are removed from the
%graph. Whenever a check node has only a single neighbor bit node
%left, the value of the corresponding bit is recovered from the value
%of the check node, and the removing process is repeated.

The minimum-delay decoding of non binary codes consists of removing
symbols from the sets of eligible symbols:
\begin{itemize}
  \item initialize $\msr{E}_n = \gf_q$, $n=1,\dots,N$
  \item each time a new bit is received, identify the variable node $n$ of which the received bit is a constituent bit, and then:
  \begin{description}
    \item[A($n$):] remove symbols from $\msr{E}_n$ whose corresponding constituent bit is different from the received bit
    \item[B($n$):] process the check nodes $m\in\mcl{H}(n)$, then update the sets of eligible symbols $\msr{E}_{n'} \leftarrow \overline{\msr{E}}_{n'}$, for each $n'\in
    \mcl{H}(m)\setminus\{n\}$
    \item[C($n$):] For each of the above $n'$s, if by updating $\msr{E}_{n'}$ its cardinal has been reduced, go to B($n\leftarrow n'$).
  \end{description}
\end{itemize}
The decoder stops when all the sets $\msr{E}_n$ are of cardinality
$1$.

\subsubsection{Decoding inefficiency}
 It follows that the minimum-delay decoding is actually an {\em on-the-fly} implementation of the previous iterative decoding.
A performance metric that is often associated with on-the-fly
decoding is the decoding inefficiency, defined as the ratio between
the number of received bits before decoding stops and the number of
information bits. Let $K_\txtscript{bin}$ be the binary dimension of
the code, and $K_\txtscript{received}$ be the number of received
bits before decoding stops. Then the inefficiency $\mu$ is defined
as:
\begin{equation}
  \mu = \frac{K_\txtscript{received}}{K_\txtscript{bin}}
\end{equation}
The expectation of the random variable $\mu$, denoted by $\muave$, is called {\em average inefficiency}. In practice $\muave$ can be
estimated by Monte-Carlo simulation.

The average inefficiency of the on-the-fly decoding can be related
to the failure probability of the iterative decoding (section
\ref{decoding_bec}). Precisely, for any $\epsilon\in[0,1]$, let
$p(\epsilon)$ be the failure probability of the iterative decoding
assuming that $\epsilon$ is the channel erasure probability.
Assuming that the function $p$ is integrable on $[0,1]$, we have:
\begin{equation}
  \mu_m - 1 = \int_0^1 p(\epsilon) \,d\epsilon
\end{equation}

\section{Density evolution} \label{density_evolution_bec}
%In this section we use the density evolution approach to analyse the iterative decoder of non binary LDPC codes.
Density evolution for non binary LDPC codes over the BEC was already
derived in \cite{rathi:det}, assuming an uniform distribution on the
edge labels. In {\em loc. cit.}, the authors suggest that the
distribution of the edge labels represents a degree of freedom that
should be integrated to our understanding of capacity approaching
iterative coding schemes. To do so, we derive the density evolution
of non binary codes tacking into consideration the variable and
check nodes degree distributions, but also the probability
distribution of the edge labels. We use the following notation:
\begin{itemize}
\item $\lambda_d$ is the fraction of edges connected to variable nodes of degree $d$, $\lambda(X) = \displaystyle\sum_{d=1}^{\dvmax}\lambda_d
X^{d-1}$ is the polynomial of  variable node degree distribution;
\item $\rho_d$ is the fraction of edges connected to check nodes of degree $d$, $\rho(X) = \displaystyle\sum_{d=1}^{\dcmax}\rho_d
X^{d-1}$ is the polynomial of  check node degree distribution;
\item $f : \mbb{L} \rightarrow [0,1]$ the probability distribution function defined by $f(h) = $ fraction of edges with label
$h\in\mbb{L}$. By extending the notation, for a given sequence $\mbf{h} = (h_1,\dots, h_I)$ we define $f(\mbf{h}) =
\displaystyle\prod_{i=1}^{I}f(h_i)$.
\end{itemize}
Without losing generality, we may assume that the all-zero codeword
is transmitted. Thus, any set of eligible symbols ($\msr{E}_n,
\msr{A}_{m,n}, \msr{B}_{m,n}$, or $\overline{\msr{E}}_{n}$) is a
$\gf_2$-linear sub-spaces of $\gf_q$. Table \ref{E_values} gives the
list of the possible values of the a priori sets of eligible symbols
$\msr{E}_n$ for the case of a $\gf_8$-code, according to the
received binary sequence\footnote{Here we identify $\gf_8 = \{0, 1,
2,\dots, 7\}$, and the constituent bits of a given symbol correspond
to the binary decomposition.}.

\begin{table}[!h]
 \caption{Possible values of the a priori sets of eligible symbols}
 \label{E_values}
 \vspace{-5mm}
 $$\begin{array}{|c|c|c|}
    \hline
    \mbox{received bits}^{*} & \msr{E}_n & \Pr(\msr{E}_n) \\
    \hline\hline
    \mbox{x} \mbox{x} \mbox{x} & \gf_8 & \epsilon^3\\
    \hline
    0 \mbox{x} \mbox{x} & \{ 0, 1, 2, 3 \} & \epsilon^2(1-\epsilon)\\
    \hline
    \mbox{x} 0 \mbox{x} & \{ 0, 1, 4, 5 \} & \epsilon^2(1-\epsilon)\\
    \hline
    \mbox{x} \mbox{x} 0 & \{ 0, 2, 4, 6 \} & \epsilon^2(1-\epsilon)\\
    \hline
    \mbox{x} 0 0 & \{ 0, 4 \} & \epsilon(1-\epsilon)^2\\
    \hline
    0 \mbox{x} 0 & \{ 0, 2 \} & \epsilon(1-\epsilon)^2\\
    \hline
    0  0 \mbox{x} & \{ 0, 1 \} & \epsilon(1-\epsilon)^2\\
    \hline
    0  0 0 & \{ 0 \} & (1-\epsilon)^3\\
    \hline
 \end{array}$$
 \begin{center}
   ${}^{*}$ Symbol ${\mbox{x}}$ denotes an erased bit
 \end{center}
\end{table}

\vspace{-3mm}
 Let $\Gr(\gf_q)$  be the Grassmannian of $\gf_q$, that is the set of all $\gf_2$-linear subspaces of $\gf_q$. For
$V\in\Gr(\gf_q)$, we note:
\begin{eqnarray}
 P_\ell(V) & = & \Pr(\msr{A}_{m,n}^{(\ell)} = V)\\
 Q_\ell(V) & = & \Pr(\msr{B}_{m,n}^{(\ell)} = V)
\end{eqnarray}
where superscript $(\ell)$ is used to denote sets of eligible
symbols computed at the $\ell^\txtscript{th}$ iteration. Thus, the
decoding is successfully if and only if:
\begin{equation}
  \label{success_cond}
  \lim_{\ell \rightarrow +\infty} P_\ell(\{0\}) = 1
\end{equation}

In order to simplify the notation, we define:
\begin{itemize}
\item For any $V\in\Gr(\gf_q)$:
 $$\begin{array}{r@{\ }c@{\ }l}
 \gamma(V) & {:}{=} & P_0(V)  =  \Pr(\msr{E}_n = V)\\
 \mcl{S}_V^{(I)} & {:}{=} & \{ \umbf{V} = (V_1, \dots, V_I) \mid  \displaystyle\sum_{i=1}^{I}V_i = V\} \subseteq \Gr(\gf_q)^I \\
 \mcl{I}_V^{(I)} & {:}{=} & \{ (V_0, \umbf{V}) = (V_0, V_1, \dots, V_I) \mid \displaystyle\bigcap_{i=0}^{I}V_i =
 V\}\\
 & \subseteq & \Gr(\gf_q)^{I+1}
 \end{array}$$
\item For any $\mbf{h} = (h_1,\dots,h_I)\in \mbb{L}^{I}$ and $\umbf{V} = (V_1, \dots, V_I)\in
\Gr(\gf_q)^{I}$:
 $$\mbf{h}^{-1} {:}{=} (h_1^{-1}, \dots, h_I^{-1}),\ \ \mbf{h}\cdot \umbf{V} {:}{=} (h_1 V_1, \dots, h_I V_I)$$
\item For any $\umbf{V} = (V_1, \dots, V_I)\in \Gr(\gf_q)^{I}$:
 $$P_\ell(\umbf{V}) {:}{=} \prod_{i=1}^{I} P_l(V_i),\ \ Q_\ell(\umbf{V}) {:}{=} \prod_{i=1}^{I} Q_l(V_i)$$
%\item $\mcl{S}_V^{(I)} = \{ \umbf{V} = (V_1, \dots, V_I) \in \Gr(\gf_q)^I \mid  \displaystyle\sum_{i=1}^{I}V_i = V\}$
%\item $\mcl{I}_V^{(I)} = \{ (V_0, \umbf{V}) = (V_0, V_1, \dots, V_I) \in \Gr(\gf_q)^{I+1} \mid  \displaystyle\bigcap_{i=0}^{I}V_i = V\}$
\end{itemize}

Let $(m,n)$ be an edge of the tanner graph. Assume that
$\mcl{H}(m)=\{n,n_1,\dots,n_{d-1}\}$, where $d$ is the degree of the
check node $m$. To simplify the notation, we set $h_i = h_{m,n_i}$,
the non zero label of the edge $(m,n_i)$, for $i=1,\dots, d-1$. The
probability of $\msr{B}_{m,n}^{(\ell+1)}$ being equal to $V$,
conditioned on $\mbf{h} = (h_1,\dots,h_{d-1})$, may be computed as:
\begin{equation}
\Pr(\msr{B}_{m,n}^{(\ell+1)} = V\mid \mbf{h}) = \sum_{\umbf{V} \in \mcl{S}_V^{(d-1)}} \left(\ \prod_{i=1}^{d-1} P_\ell(h_i^{-1}V_i)\
\right)
\end{equation}
Averaging over all possible label sequences $\mbf{h}$ we get:
\begin{eqnarray}
 Q_{\ell+1}^{(d-1)}(V) \hspace{-2mm}& {:}{=} \hspace{-2mm}& \Pr(\msr{B}_{m,n}^{(\ell+1)} = V) \nonumber \\
  & = \hspace{-4mm}& \sum_{\mbf{h}\in\mbb{L}^{d-1}}\left(f(\mbf{h})\cdot\!\!\!\sum_{\umbf{V} \in
\mcl{S}_V^{(d-1)}} \!\!\!\!\!P_\ell(\mbf{h}^{-1}\cdot\umbf{V})
\right)
\end{eqnarray}
Averaging over all possible check node degrees $d$, we obtain:
\begin{equation}
Q_{\ell+1}(V) = \sum_{d=1}^{\dcmax} \left(\rho_d \cdot Q_{\ell+1}^{(d-1)}(V)\right)
%\!\!\!\!\sum_{\mbf{h}\in(\gf_q^{*})^{d-1}}\left(f(\mbf{h})\cdot\!\!\!\sum_{\umbf{V} \in \mcl{S}_V^{(d-1)}}
%\!\!\!\!\!P_\ell(\mbf{h}^{-1}\cdot\umbf{V}) \right)
\end{equation}

Now, consider an edge $(n,m)$ of the Tanner graph, and let the
variable node $n$ be of degree $d$ and
$\mcl{H}(n)=\{m,m_1,\dots,m_{d-1}\}$. To simplify notation, we set
$h_i = h_{m_i,n}$, the non zero label of the edge $(n, m_i)$, for
$i=1,\dots, d-1$. The probability of $\msr{A}_{m,n}^{(\ell+1)}$
being equal to $V$, conditioned on $\mbf{h} = (h_1,\dots,h_{d-1})$,
may be computed as:
\begin{equation}
\Pr(\msr{A}_{m,n}^{(\ell+1)} = V\mid \mbf{h}) =
\hspace{-4mm}\sum_{(V_0,\umbf{V}) \in \mcl{I}_V^{(d-1)}}
\hspace{-1mm}\left(\ \gamma(V_0)\prod_{i=1}^{d-1} Q_{\ell+1}(h_i
V_i)\ \right)
\end{equation}
Again, by averaging over all possible label sequences $\mbf{h}$, it
follows that:
\begin{eqnarray}
 \lefteqn{P_{\ell+1}^{(d-1)}(V)\ {:}{=}\ \Pr(\msr{A}_{m,n}^{(\ell+1)} = V)} \nonumber \\
 & = & \sum_{\mbf{h}\in\mbb{L}^{d-1}}\hspace{-1mm}\left(f(\mbf{h})\cdot\hspace{-3mm}\sum_{(V_0,\umbf{V}) \in
\mcl{I}_V^{(d-1)}} \!\!\!\!\!\gamma(V_0) Q_{\ell+1}(\mbf{h}\cdot
\umbf{V})\right)
\end{eqnarray}
Finally, averaging over all possible variable node degrees $d$, we
obtain:
\begin{equation}
P_{\ell+1}(V) = \sum_{d=1}^{\dvmax} \left(\lambda_d \cdot P_{\ell+1}^{(d-1)}(V)\right)
%\!\!\!\!\sum_{\mbf{h}\in(\gf_q^{*})^{d-1}}\left(f(\mbf{h})\cdot\!\!\!\sum_{(V_0,\umbf{V}) \in \mcl{I}_V^{(d-1)}} \!\!\!\!\!\gamma(V_0)
%Q_{\ell+1}(\mbf{h}\cdot \umbf{V})\right)
\end{equation}

\begin{prop}
  Let $V, W\in \Gr(\gf_q)$ and $h\in\mbb{L}$ such that $W = hV$. Then:
  \begin{eqnarray}
    Q_{\ell+1}^{(d-1)}(W)\hspace{-3mm}
    & = \hspace{-3mm}& \hspace{-2mm}\sum_{\mbf{h}\in\mbb{L}^{d-1}}\hspace{-1mm}\left(f(h\cdot\mbf{h})\cdot\hspace{-1mm}\!\!\!\sum_{\umbf{V} \in
    \mcl{S}_V^{(d-1)}} \!\!\!\!\!P_\ell(\mbf{h}^{-1}\cdot\umbf{V}) \right) \\
    P_{\ell+1}^{(d-1)}(W) \hspace{-3mm}
    & = \hspace{-3mm}& \nonumber \\
    \lefteqn{\hspace{-12mm}\sum_{\mbf{h}\in\mbb{L}^{d-1}}\left(f(h\cdot\mbf{h})\cdot\!\!\!\sum_{(V_0,\umbf{V}) \in
    \mcl{I}_V^{(d-1)}} \!\!\!\!\!\gamma(V_0) Q_{\ell+1}(\mbf{h}\cdot
    \umbf{V})\right)}
  \end{eqnarray}
where $h\cdot(h_1,\dots,h_{d-1}) = (h h_1,\dots,h h_{d-1})$. In particular, if $f$ is the uniform distribution, then $Q_{\ell+1}^{(d-1)}(W)
= Q_{\ell+1}^{(d-1)}(V)$ and $P_{\ell+1}^{(d-1)}(W) = P_{\ell+1}^{(d-1)}(V)$.
\end{prop}

We say that $V$ and $W$ are conjugate if there exists $h\in\mbb{L}$ such that $W = hV$ and denote by $\Gr(\gf_q)/\mbb{L}$ the quotient set
of conjugation classes.

\begin{coro}
  Assume that $f$ is the uniform distribution and let $V\in\Gr(\gf_q)$. Then $Q_{\ell}(V)$ and $P_{\ell}(V)$ depend
  only on the conjugation class of $V$ in $\Gr(\gf_q)/\mbb{L}$.
\end{coro}

\begin{coro}
  Assume that $f$ is the uniform distribution and that $\mathbb{L} = \mbf{M}_p^{*}(\gf_2)$, the multiplicative group of invertible $p\times p$ matrices,
  acting on $\gf_q$ via the isomorphism $\gf_2^p \stackrel{\sim}{\rightarrow} \gf_q$ from (\ref{identify}).  Let $V\in\Gr(\gf_q)$. Then $Q_{\ell}(V)$
  and $P_{\ell}(V)$ depend only on the dimension of the vector space $V$.
\end{coro}

The above corollaries may be used to simplify the density evolution
formulas, assuming a uniform distribution of the graph edge labels.
For instance, if $\mathbb{L} = \mbf{M}_p^{*}(\gf_2)$, one can derive
the same formulas as in \cite{rathi:det}.

\section{Thresholds}\label{thresholds}
We denote by $E_{\gf_q,\mbb{L}}(\lambda,\rho, f)$ the ensemble of
LDPC codes over $\gf_q$, with labels group $\mbb{L}$, distribution
degree polynomials $\lambda$ and $\rho$, and probability
distribution of edge labels $f$. Whenever the Galois group $\gf_q$
and the labels group $\mbb{L}$ are subunderstood, we will simply use
$E(\lambda,\rho, f)$. We also denote by
${\pth}_{\gf_q,\mbb{L}}(\lambda,\rho, f)$ (or simply
$\pth(\lambda,\rho, f)$) the threshold probability of the above
ensemble, that is (see also (\ref{success_cond})):

\vspace{-5mm}
\begin{equation}
  \vspace{-2mm}
  \pth(\lambda,\rho, f) = \max \{ \epsilon \mid \lim_{\ell \rightarrow +\infty}\!\!\! P_\ell(\{0\}) = 1 \mbox{ on BEC}(\epsilon) \}
  %\mbox{ where } P_\ell = \sum_{\stackrel{V \in \Gr(\gf_q)}{V \varsupsetneq\{0\}}} P_{\ell}(V)
\end{equation}

\begin{figure}[t!]
  \begin{center}
    \includegraphics[width=0.8\linewidth]{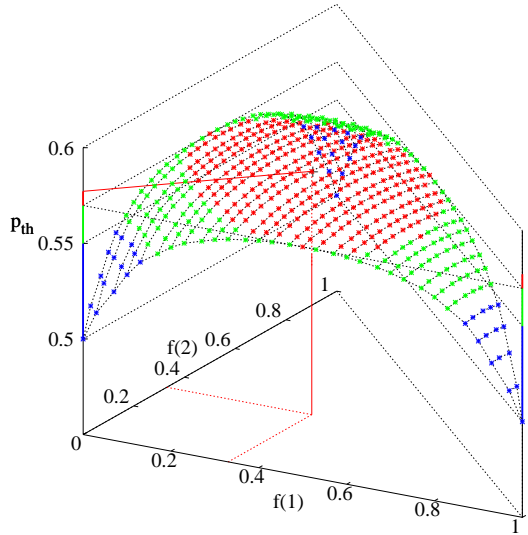}
  \end{center}
  \caption{Probability threshold of the ensemble $E_{\gf_4,\gf_4^{*}}(\lambda = X,\rho = X^2, f)$ as function of labels probability distribution
  $f$.}
  %Edge labels are sampled from $\gf_4^{*} \cong \{1,2,3\}$ according to the probability distribution $f$.
  %The $x$-axis $f(1)$ represents
  %the probability of the edge label being $1$, and the $y$-axis $f(2)$ represents the probability of the edge label being $2$
  %(thus the probability of the edge label being $3$ is given by $f(3) = 1 - f(1) - f(2)$).}
  \label{courbe_thres_gf4_max}
\end{figure}

By fixing the polynomials of degree distribution $\lambda$ and
$\rho$, the probability threshold $\pth$ may be seen as a function
of the probability distribution $f$. This is illustrated in Fig.
\ref{courbe_thres_gf4_max} and Fig. \ref{courbe_thres_gf4_min}. The
Galois field is $\gf_4$ and the labels group $\mbb{L} = \gf_4^{*}$,
acting of $\gf_4$ by the internal field multiplication. The
horizontal axes $f(1)$ and $f(2)$ represent the probabilities of
edge labels being $1$ and $2$, respectively. Thus, the probability
of edge labels being $3$ is given by $f(3) = 1 - f(1) - f(2)$. We
drawn the surface representing $\pth$ as function of $f(1)$ and
$f(2)$. The top of the surface is plotted in red, the middle in
green, and the bottom in blue. The two figures correspond to two
couples $(\lambda,\rho)$ of degree distributions that were also
considered in \cite{rathi:det}. In Fig. \ref{courbe_thres_gf4_max}
we fix $\lambda=X$ and $\rho=X^2$. The maximum $\pth$ is obtained
for the uniform distribution $f(1) = f(2) = f(3) = 1/3$ and its
value is equal to $0.5772$. The minimum $\pth = 0.5$ is obtained for
the three distributions concentrated in a single label (such codes
are equivalent to binary codes). In Fig. \ref{courbe_thres_gf4_min}
we fix $\lambda(X) = X^2$ and $\rho(X) = X^3$. For the uniform
distribution $f(1) = f(2) = f(3) = 1/3$, the threshold $\pth =
0.6348$. The minimum $\pth = 0.6346$. The maximum $\pth = 0.6474$ is
obtained for the three distributions concentrated in one single
label.

These two examples highlight a more general phenomenon that we
observed for other ensembles of codes, as shown for instance in Fig.
\ref{courbe_thres_gf4_gf8}. For a given Galois field $\gf_q$, and
given polynomials  $\lambda$, and $\rho$, it is possible to find a
probability distribution $\tilde{f}$ of edge labels, such that:
\begin{itemize}
  \item edge labels are equal to $1$ with high probability
(meaning that $\tilde{f}(1)$ is close to $1$)
  \item ${\pth}_{\gf_q,\gf_q^{*}}(\lambda,\rho, \tilde{f}) \thickapprox \displaystyle \max_f {\pth}_{\gf_q,\gf_q^{*}}(\lambda,\rho, {f})$
\end{itemize}

\addtocounter{figure}{1}
\begin{figure}[h!]
  \begin{minipage}{.93\linewidth}
  \raggedleft
  \footnotesize
  \begin{tabular}{c@{\,}r}
  \rotatebox{90}{\hspace{-13mm}$E_{\gf_8,\gf_8^{*}}(\lambda,\rho, f)$} &
  \begin{tabular}{|*{7}{c|}@{}c@{\,\,}|c|}
    \hline
    \multicolumn{7}{|c|}{Labels pdf $f$} && Threshold \\
    \cline{1-7}
    1 & 2 & 3 & 4 & 5 & 6 & 7 &&  \\
    \hline \hline
    1/7 & 1/7 & 1/7 & 1/7 & 1/7 & 1/7 & 1/7 && 0.4353 \\
    \cline{1-7}\cline{9-9}
    1/5 & 1/5 & 1/5 & 0 & 0 & 1/5 & 1/5 && 0.4356 \\
    \cline{1-7}\cline{9-9}
    1/3 & 0 & 0 & 1/3 & 0 & 0 & 1/3 && 0.4373 \\
    \cline{1-7}\cline{9-9}
    1/3 & 1/3 & 1/3 & 0 & 0 & 0 & 0 && 0.4391\\
    \cline{1-7}\cline{9-9}
    1/2 & 0 & 0 & 0 & 0 & 0 & 1/2 && 0.4437 \\
    \cline{1-7}\cline{9-9}
    0.8 & 0 & 0 & 0 & 0 & 0 & 0.2 && 0.4483\\
    \cline{1-7}\cline{9-9}
    0.9 & 0 & 0 & 0 & 0 & 0 & 0.1 && 0.436 \\
    \cline{1-7}\cline{9-9}
    0.95 & 0 & 0 & 0 & 0 & 0 & 0.05 && 0.4179\\
    \cline{1-7}\cline{9-9}
    1 & 0 & 0 & 0 & 0 & 0 & 0 && 0.4\\
    \hline
    %\rotatebox{90}{$E_{\gf_8,\gf_8^{*}}(\lambda,\rho, f)$}
  \end{tabular}
  \end{tabular}

  \vspace{1.5mm}
  \begin{tabular}{c@{\,}r}
  \rotatebox{90}{\hspace{-12mm}$E_{\gf_4,\gf_4^{*}}(\lambda,\rho, f)$} &
  \begin{tabular}{|*{3}{c|}@{}c@{\,\,}|c|}
    \hline
    \multicolumn{3}{|c|}{Labels pdf $f$} && Threshold \\
    \cline{1-3}
    1 & 2 & 3  &&  \\
    \hline \hline
    1/3 & 1/3 & 1/3 && 0.4487 \\
    \cline{1-3}\cline{5-5}
    1/2 & 1/2 & 0 && 0.4489 \\
    \cline{1-3}\cline{5-5}
    0.8 & 0.1 & 0.1 && 0.4507 \\
    \cline{1-3}\cline{5-5}
    0.9 & 0.07 & 0.03 && 0.4335 \\
    \cline{1-3}\cline{5-5}
    0.97 & 0.03 & 0 && 0.4121 \\
    \cline{1-3}\cline{5-5}
    1 & 0 & 0 && 0.4\\
    \hline
  \end{tabular}
  \end{tabular}
  \end{minipage}
  \caption{Probability thresholds of the ensembles $E_{\gf_8,\gf_8^{*}}(\lambda,\rho, f)$ and $E_{\gf_4,\gf_4^{*}}(\lambda,\rho, f)$
  for $\lambda = 0.5 X + 0.5 X^4$, $\rho = X^5$, and different labels probability
  distributions $f$.}
  \label{courbe_thres_gf4_gf8}
  \vspace{-5mm}

\end{figure}

For instance, considering the ensemble of codes over $\gf_8$ from
Fig. \ref{courbe_thres_gf4_gf8}, if $\tilde{f}$ is defined by
$\tilde{f}(1) = 0.8$, $\tilde{f}(7) = 0.2$, and $\tilde{f}(i) = 0$
for $1 < i < 7$, then ${\pth}_{\gf_8,\gf_8^{*}}(\lambda,\rho,
\tilde{f}) = 0.4483$. In this case only few Galois field
multiplications are needed, and the decoder complexity is
considerably reduced.

\addtocounter{figure}{-2}
\begin{figure}[t!]
  \begin{center}
    \includegraphics[width=0.8\linewidth]{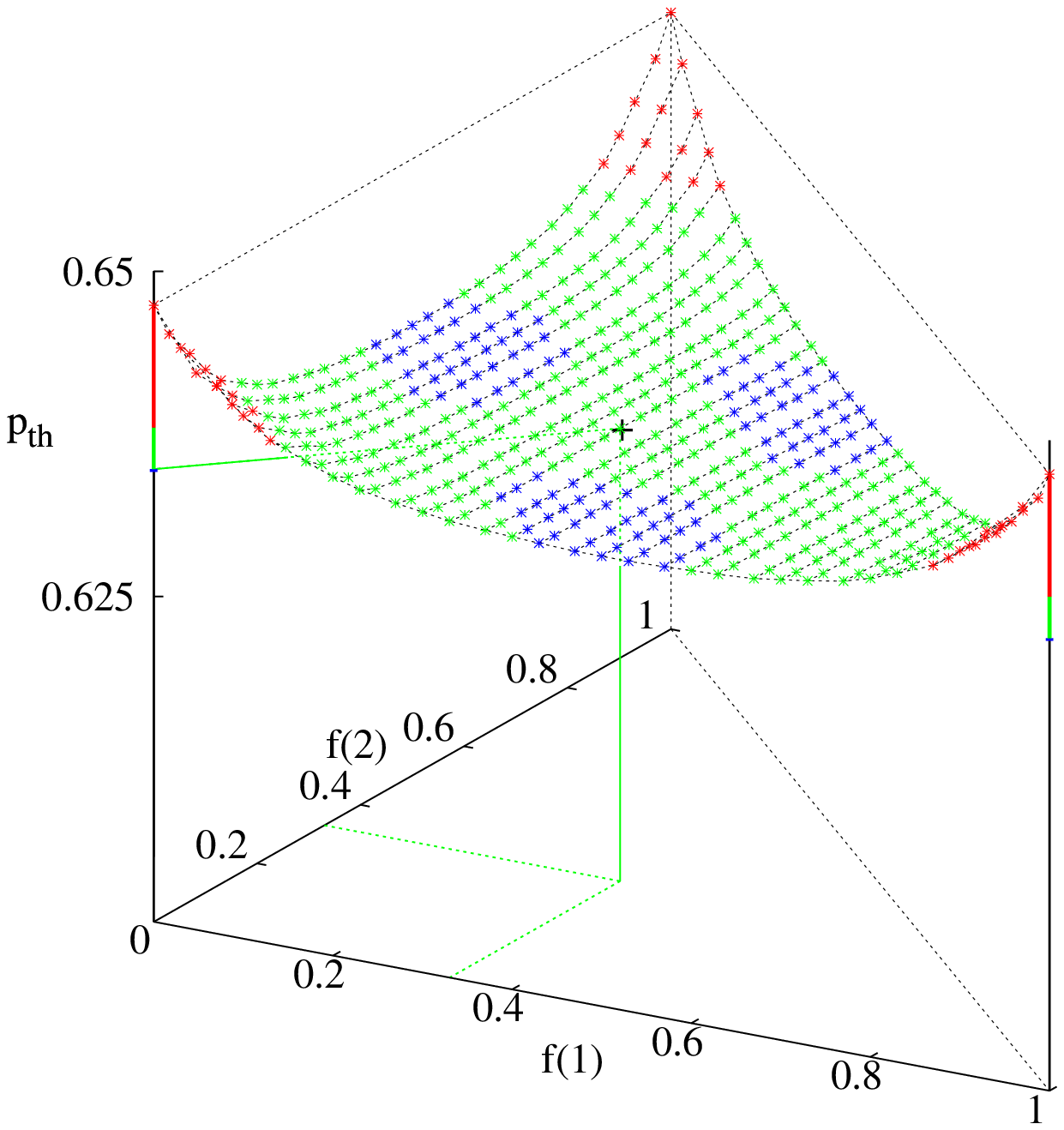}
  \end{center}
  \caption{Probability threshold of the ensemble $E_{\gf_4,\gf_4^{*}}(\lambda = X^2,\rho = X^3, f)$ as function of labels probability distribution
  $f$.}
  %Edge labels are sampled from $\gf_4^{*} \cong \{1,2,3\}$ according to the probability distribution $f$. The $x$-axis $f(1)$ represents
  %the probability of the edge label being $1$, and the $y$-axis $f(2)$ represents the probability of the edge label being $2$
  %(thus the probability of the edge label being $3$ is given by $f(3) = 1 - f(1) - f(2)$).}
  \label{courbe_thres_gf4_min}
  \vspace{-5mm}

\end{figure}

\section{Conclusions}
In this paper we investigated the decoding of non binary LDPC codes
over the BEC, and we introduced a minimum-delay decoding suited for
UL-FEC. We also derived the density evolution equations taking into
consideration both the irregularity of the bipartite graph of the
code and the probability distribution of the graph edge labels,
giving a thorough understanding of the asymptotical behavior of
ensembles of non binary LDPC codes. A non-uniform probability
distribution of the edge labels might improve the decoder
performance, but the most important advantage is that the decoder
complexity can be significantly reduced. The design of capacity
approaching non binary LDPC codes will be addressed in future works.

 \bibliographystyle{./bib/IEEEbib}
\footnotesize
\bibliography{./bib/MyBiblio,./bib/Zotero}

\end{document}